\documentclass[twocolumn,showpacs]{revtex4}%
\usepackage{graphicx}
\usepackage{amsmath}
\usepackage{amsfonts}
\usepackage{amssymb}
\usepackage{color}
\usepackage{amscd}
\usepackage{bm}
\usepackage{slashbox}
\usepackage{float}

\begin{document}
\title{The utmost distance for quantum entanglement}
\author{Yong Xiao}
\email{xiaoyong@hbu.edu.cn}
\address{College of Physical Science and Technology, Hebei University, Baoding 071002, China}

\begin{abstract}
A common viewpoint is that a particle could be quantum entangled
with another particle arbitrarily far away. But in this paper we
suggest that there is an utmost distance for the existence of
quantum entanglement between two particles, beyond which the initial
quantum entanglement would be broken by some quantum gravitational
effect. The utmost distance is proposed to be $L_{QE}=\lambda^\alpha
l_{p}^{1-\alpha}$, where $\lambda$ is the quantum wavelength of the
particles and $l_{p}= 1.616 \times 10^{-35} m$ is the Planck length.
The most probable value of the parameter $\alpha$ is $2$ or $3$. As
other quantum-gravitational effects, this effect is very weak and
hard to be detected in foreseeable experiment.

\end{abstract}
\maketitle

\address{College of Physical Science and Technology, Hebei University, Baoding 071002, China}

\section{Introduction}

Quantum entanglement (QE) is one of the most mysterious phenomena of
quantum mechanics, well known as \textquotedblleft spooky action at
a distance". When measuring one of the two entangled particles, the
other particle acquires the corresponding state simultaneously.
Nevertheless, due to the uncontrollable randomness, the existence of
quantum no-cloning theorem and other subtleties in the process, the
phenomenon cannot be used to transmit classic information in a
superluminal way. So it avoids a direct confliction between quantum
mechanics and special relativity.

An immediately question is how far the QE between two particles
could be retained. The common answer is that two particles can be
entangled at an infinite distance. This is understandable in the
framework of quantum mechanics and special relativity, because one
can never construct a fundamental distance from the reduced Planck
constant $\hbar$, the speed of light $c$, and the quantum wavelength
$\lambda$ characterizing the particle. The absence of a candidate
distant leads to the acceptance of an infinite distance for QE,
though the emergence of infinity usually implies our ignorance about
the relevant physics (the boundary between quantum and classical
physics in this case). Experimental physicists are willing to judge
the foundations of quantum physics by direct experiment. Recently,
the quantum satellite ``Micius'' successfully distributed entangled
pairs of photons to separate stations $1200$ kilometers apart and
proved that QE still exists in such a remote distance \cite{pan}.
The team further plans to send a photon source to the Lagrange point
of Earth-moon system and to test the robustness of QE at a distant
over $300,000$ kilometers. It is conceivable that this kind of
experiment over such an ultra-large scale may challenge our
understanding of gravity and the structure of spacetime.

Actually when gravitational effects are included, the theoretical
analysis before has to be changed. After introducing the
gravitational constant $G$ into the analysis, a fundamental length
known as Planck length can be constructed, which is
$l_{p}=\sqrt{\hbar G/c^{3}}= 1.616 \times 10^{-35} m $. Then
combining with the characteristic wavelength $\lambda$ of the
entangled particles, we can construct a length
$L_{QE}=\lambda^\alpha l_{p}^{1-\alpha}$ by dimensional analysis,
where $\alpha$ is a parameter to be analyzed below. If there is
really an utmost distance for QE, it naturally has this form and the
only question is how to choose the value of $\alpha$. In the
following, we shall illustrate that the most probable value of
$\alpha$ is $2$ or $3$.

\section{The reason for $\alpha=2$}

Quantum mechanics is the low energy approximation of quantum field
theory (QFT). The microscopic particles such as photons, electrons
are the excitations of quantum fields. It is realized that QFT is
incomplete at ultra-high energy and ultra-large length scale. At
high energy scale, QFT has various infinities to be renormalized and
for ultra-high energy the self-gravitation of the quantum fields is
too strong to be omitted, while a rigorous description of quantum
gravity is still absent. At super-large length scale, the vacuum
energy of conventional QFT has an enormous discrepancy with the
observed cosmological constant.

In \cite{cohen}, Cohen et al. pointed out that conventional QFT
cannot be correct at a rather large length scale, after
gravitational limitations have been taken into consideration. They
suggested a relation $L\leq m_{p}/\Lambda^{2}$ between the
ultraviolet (UV) cutoff $\Lambda$ and infrared (IR) cutoff $\hbar/L$
as the applicable condition for the conventional QFT and argued that
it doesn't conflict with the present particle physics experiment. As
an example, when the UV-IR relation is applied to the size of
cosmological horizon, the corresponding UV cutoff is
$\Lambda\sim10^{-2.5}eV$ and eliminates the necessity of fine-tuning
of the vacuum energy.

Since the conventional QFT ceases to be in effect for $\Lambda>
10^{-2.5}eV$ at cosmological size, QE as a fundamental concept of
quantum theory also becomes questionable. In this spirit, we may
translate the above UV-IR relation to be $L_{QE}\leq \lambda^2/
l_{p}$ and view it as the applicable condition of the concept of QE.
If the conjecture is correct, the particles with momentum
$\Lambda\leq 10^{-2.5}eV$ (wavelength $\lambda\geq 4\times
10^{-4}m$) can be entangled with each other at the cosmological
size, and for the particles with higher momentum (shorter
wavelength), the QE at the cosmological size may be broken by some
quantum gravitational ``noises'' beyond conventional QFT.

We conjecture the formula of $L_{QE}$ is applied not only to the QE
between the space configurations of the particles, such as the case
$|\phi(\vec{r}_{1})\psi(\vec{r}_{2})\rangle\pm|\phi(\vec{r}_{2})\psi(\vec{r}_{1})\rangle$,
but also to the QE between particle spins or polarizations. This is
because particle spin is a concept highly correlated with the space
configuration of the quantum fields. In the viewpoint of QFT,
particle spin is a property of quantum state under coordinate
transformations, and technically speaking the photon fields and
electron fields are respectively the vector and spinor
representations of Lorentz group. Another concrete example is that
the entire wavefunction of two electrons is anti-symmetric, so when
the space configuration part of the wavefunction is anti-symmetric,
the spin part of the wavefunction must be symmetric, and vice versa.

\section{The reason for $\alpha=3$}

Since the introduction of entropy and the second law of
thermodynamics, the concept of information has become more and more
important in physics. In recent years, inspired by black hole
physics and the generalized second law of thermodynamics, it is
realized that the maximum realizable entropy of a system of size $L$
is its boundary area in Planck units, i.e., $A/l_{p}^{2}$. This is
called holographic principle \cite{hooft,susskind,bousso}, because
it strongly implies that all the information inside a
quantum-gravitational system can be encoded on the boundary of the
system. The area-scaling entropy is often called holographic
entropy. Cohen et al. have also considered another UV-IR relation
$L\leq m_p^2/\Lambda^3 $ based on the holographic principle of
quantum gravity. As always, we translate it as a limitation to the
distance of QE, which is $L_{QE}\leq \lambda^3/ l_{p}^2$.

In fact, we prefer $L_{QE}\leq \lambda^3/ l_{p}^2$ to $L_{QE}\leq
\lambda^2/ l_{p}$ as the utmost distance for QE, because the UV-IR
relation $L\leq m_p^2/\Lambda^3 $ is derived from the holographic
principle. Holographic principle was originally extracted from the
analysis of space-time horizon which shields all the information
behind it and finally it sets a limitation to the maximum
information capacity of a general system. Note that QE is also a
concept intimately related to information. From this perspective,
 QE is conceptually close to holographic principle. In contrast, the
UV-IR relation  $L\leq m_p/\Lambda^2 $ comes from the energy
constraint to a conventional QFT system. In this kind of situations
where other limitations are tighter than the holographic principle,
because it doesn't touch the fundamental information bound, whether
the entanglement distance is finite or infinite usually becomes an
irrelevant question. (It doesn't mean there is no limitation to
entanglement distance, it only means one cannot touch it in these
situations.)

As a fundamental principle, holographic principle must take effect
independently at some physical situation where too much information
and mass are accumulated in a region. So hereafter we only consider
the limitation from holographic principle, and release other
possible quantum-gravitational limitations \cite{note}.

Now we start to provide an informational perspective argument. The
reader, who is unfamiliar with discussions about quantum gravity,
only needs to memorize that the maximum entropy inside a region is
$A/l_{p}^{2}$ where $A$ is the boundary area of the system. We show
clearly the reason why an infinite entanglement distance conflicts
with the holographic entropy limitation and how the requirement
$L_{QE}\leq \lambda^3/ l_{p}^2$ is consistent with the holographic
principle.

\begin{figure}[t]
\includegraphics[width=7cm]{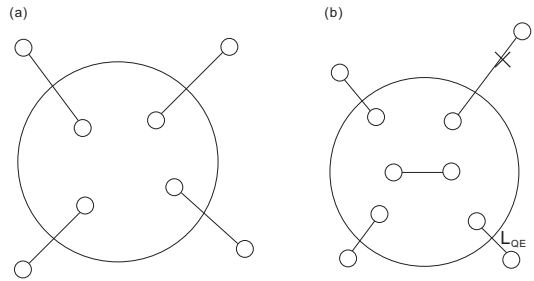}\caption{$(a)$ The entangled pairs are described by the quantum
state $|\psi\rangle=|10\rangle\pm|01\rangle$. Assume the QE can
exist at an infinite distance.  Accumulate particles inside a box of
radius $R$ and leave the particles entangled with them outside. We
can accumulate at most $N=R^3/\lambda^{3}$ particles inside the box,
where $\lambda$ is the quantum wavelength of the particles.
Measuring the box, the corresponding Von Neumann entropy is $S=N\ln
2$. The holographic principle is violated when $\lambda$ is short
enough. $(b)$ Require the utmost distance for QE is $L_{QE}=
\lambda^3/ l_{p}^2$. For the case $\lambda>R^{1/3}l_p^{2/3}$,
$L_{QE}>R$ and the entropy is $S=N\ln 2 < A/l_{p}^{2}\ln 2$ causing
no trouble. For the case $\lambda < R^{1/3}l_p^{2/3}$, we have
$L_{QE} < R$. The result is that only a layer of the particles close
to the boundary of the box can contribute to the entropy of the
system. The effective number of the particles is $N_{eff} =\frac{R^2
L_{QE} }{\lambda^3}$. Using $L_{QE}= \lambda^3/ l_{p}^2$, the
corresponding entropy is $S= A/l_{p}^{2}\ln 2$ independent of
$\lambda$.}
\end{figure}

First, we assume that QE could exist at an infinite distance.
Consider a pairs of entangled particles described by the quantum
state $|\psi\rangle=|10\rangle\pm|01\rangle$. When we observe one of
the two particles alone, we get a mixed-state density matrix and the
corresponding uncertainty in the measurement can be calculated by
the Von Neumann entropy, which gives $S=\ln2$. Now we collect many
pairs of entangled particles. Accumulate one particle of each
entangled pair in a box of radius $R$ and put the other particles
far away. We can accumulate at most $N=R^{3}/\lambda^{3}$ particles
into the box, where $\lambda$ is the quantum wavelength of the
particles. Measuring the state of the box , the corresponding Von
Neumann entropy is $S=N \ln2$. For particles with wavelength short
enough, the box has more entropy than $A/l_{p}^{2}$, which conflicts
with holographic principle.

Second, we require the utmost distance of QE to be $L_{QE}=
\lambda^3/ l_{p}^2$. Accumulate $N=R^{3}/\lambda^{3}$ particles
inside the box. For the simple case $\lambda>R^{1/3}l_p^{2/3}$,
there is $L_{QE}=\lambda^3/ l_{p}^2>R$, thus the QE of all the
entangled pairs can be retained. The Von Neumann entropy of the box
is $S=N \ln 2 < A/l_{p}^{2}\ln 2$ causing no trouble. The intriguing
consideration is about the case with $\lambda < R^{1/3}l_p^{2/3}$.
In this case $N=R^{3}/\lambda^{3}> A/l_{p}^{2}$. However, we have
$L_{QE}= \lambda^3/ l_{p}^2 < R$ meanwhile, thus many particles
cannot contribute to the calculation of the entropy. As visualized
in Fig.1b, if two particles are far apart, the QE between them has
been destroyed and the quantum state has collapsed to a certain
non-entangled state, thus the particle inside the box doesn't
contribute to the total entropy.  Another visualized situation in
Fig.1b is that both the entangled particles are inside the box to
retain the QE, the particles as a whole provides a pure state when
measuring the box and no entropy is contributed. As a result, only a
layer of the particles close to the boundary of the box can
contribute to the entropy of the system, the effective number of the
particles is $N_{eff}=N \frac{R^2 L_{QE}}{R^3}=\frac{R^2
L_{QE}}{\lambda^3}$. The corresponding Von Neumann entropy is
$S=\frac{R^2 L_{QE}}{\lambda^3}\ln 2$. Obviously, using $L_{QE}=
\lambda^3/ l_{p}^2$, we obtain the expected holographic entropy
$S=A/l_{p}^{2}\ln 2$. And the entropy cannot increase even using
particles with much shorter wavelength. Accordingly, this scenario
is nicely consistent with holographic principle.

\section{Conclusions}

In this paper, we argue that the utmost distance for QE is
$L_{QE}=\lambda^\alpha l_{p}^{1-\alpha}$, where $\lambda$ is the
quantum wavelength of the entangled particles and $l_{p}$ is the
Planck length. The possible value of $\alpha$ is $2$ or $3$. The
value $\alpha=2$ is derived from the applicable condition of
conventional QFT, and $\alpha=3$ comes from the constraints of
holographic principle of quantum gravity which limits the maximum
entropy or information inside a system.

As various quantum-gravitational effects in the literature, the
effect is rather weak and hard to be detected in experiment. As
displayed in Table I, for the entangled photon pairs with wavelength
$800nm$ employed in the quantum satellite, QE can exist over a
distance $10^{22}m$ or even $10^{51}m $.  In order to detect the
breakdown of QE at the distance $1200km$, one must use the photons
or other microscopic particles with energy up to $10Mev$ or even
$100 Tev$. So it seems very hard to find the signal of this effect
in foreseeable future.

\begin{table}[H]
\newcommand{\tabincell}[2]{\begin{tabular}{@{}#1@{}}#2\end{tabular}}
\begin{tabular}{|c|c||c|c|}
\hline \hline
\backslashbox & $\lambda=8\times 10^{-7}m$ & $L_{QE}=1.2\times 10^6 m$ & $L_{QE}=3\times 10^8 m$  \\
\hline
$\alpha=2$ &  $L_{QE}\leq 10^{22}m$ &  \tabincell{c}{$\lambda\geq 10^{-15}m$ \\ $(10 Mev)$ }     & \tabincell{c}{$\lambda\geq 10^{-14}m  $ \\ $(1 Mev)$ }  \\
\hline
$\alpha=3$ & $L_{QE}\leq 10^{51}m $ &  \tabincell{c}{$\lambda\geq 10^{-22}m$ \\ $(100 Tev)$}    & \tabincell{c}{$\lambda\geq 10^{-21}m  $ \\ $(10 Tev)$} \\
 \hline
\end{tabular}
\caption{The existence condition of QE }
\end{table}

Conventional QFT should be modified at high energy scale and large
length scale. Experimental physicists have tried to examine the
validity of conventional QFT and to find the possible quantum
gravitational effects at high energy scale in particle accelerators
for many years. And now, by examining the properties of QE, the
quantum satellite experiment and other analogous experiments provide
another opportunity to study quantum gravitational effects at large
length scales.

\section*{Acknowledgements}
I thank Y.S. Song, G.Z. Ning and Y.J Zhang for useful discussions. The work is
supported in part by the NNSF of China with Grant No. 11205045, the NSF of
Hebei province with Grant No. A2015201212 and the NSF of Hebei Eduction
Department with Grant No. YQ2014032.


\begin{thebibliography}{9}
\bibitem{note}The fact is that at present very little knowledge is known
in a totally quantum-gravational case expect for the holgographic
principle. The concept of energy and other relavant concepts should
be reconsidered in this situation, and the familiar QFT has become
invaid while a complete quantum-gravitational theory is absent. For
example, because of the energy constraint, the conventional QFT
forbids to accumulate more than $N=(A/l_{p}^2)^{3/4}$ photons inside
a box of size $L$. Hoewever, if the energy of these photons are
fully gravitational redshifted, many more photons can be accumulated
into the system. Besides, black hole can be considered to be
consisted of $N=A/l_{p}^2$ elementary constitutes with each having
the wavelength $L^{1/3}l_p^{2/3}$ and energy $hc/L$ \cite{mine,wei}.
Thus the $N=(A/l_{p}^2)^{3/4}$ limitation is not a fundamental
limitation of nature, it only means the conventional QFT cannot
provide a valid description beyond this scope. So we release this
limitation in the follwoing analysis. By comparison, we always
assume basic information concepts be applied undoutably throughout
the paper.
\end{thebibliography}
\end{document}